\pdfoutput=1
\documentclass[]{spie}  

\usepackage{graphicx}
\usepackage{color}

\usepackage{amsmath,amsfonts,amssymb}
\usepackage{bm}
\usepackage[colorlinks=true, allcolors=blue]{hyperref}

\usepackage{geometry}

\graphicspath{{./fig/}}

\newcommand{\Image}{\mathrm{Image}\!\,\,}

\title{Anti-windup PI controller for millimeter-wave adaptive optics: a Nobeyama 45m \\ radio telescope simulation}

\author[a]{Ichiro~Jikuya}
\author[b]{Yoichi~Tamura}
\author[c]{Akio~Taniguchi}
\author[b]{Masaki~Sakakibara}
\author[b]{Akinobu~Miyake}
\author[b]{Masato~Hagimoto}
\author[b]{Kianhong~Lee}
\author[b]{Chihiro~Imamura}
\author[d]{Shion~Takeno}
\author[d]{Sachiko~Okumura}
\author[e]{Nozomi~Okada}
\affil[a]{Kanazawa University, Kakuma-machi, Kanazawa, Japan}
\affil[b]{Nagoya University, Furo-cho, Chikusa-ku, Nagoya, Japan}
\affil[c]{Kitami Institute of Technology, 165 Koen-cho, Kitami, Japan}
\affil[d]{Japan Women's University, 2-8-1 Mejirodai, Bunkyo-ku, Tokyo, Japan}
\affil[e]{Osaka Metropolitan University, 3-3-138 Sugimoto Sumiyoshi-ku, Osaka, Japan}

\authorinfo{Further author information: (Send correspondence to Ichiro Jikuya)\\Ichiro Jikuya: E-mail: jikuya@se.kanazawa-u.ac.jp}

\begin{document} 
\maketitle

\begin{abstract}
This work addresses the control problem for Millimeter-wave Adaptive Optics (MAO), which we define as compensating for the distance variation between the primary reflector (M1) and the receiver. We utilize a measurement system developed by Tamura et al. to track these variations. 
The challenge is formalized as an asymptotic constant disturbance suppression problem. We demonstrate that an anti-windup proportional-integral (AWPI) controller effectively solves this problem while respecting the physical movement constraints of the optical driving system. Simulation results, based on the Nobeyama 45-m Telescope with a two-axis translating sub-reflector (M2), validate the performance of the proposed AWPI approach. 
\end{abstract}


\section{INTRODUCTION}
\label{sec:intro}  

This presentation addresses the control problem for Milliwave radio telescope Adaptive Optics (MAO), a concept redefined in the context of radio astronomy aimed at application to AtLAST\cite{Mroczkowski2025}. While conventional AO in optical/infrared telescopes compensates for atmospheric turbulence, the MAO discussed here is defined as the compensation of the distance variation between the primary reflector (M1) and the receiver. 
Tamura et al. have been developing an Excess Path Length (EPL) measurement system that radiates radio waves from a transmitter mounted on M1 to the receiver for precise metrology \cite{Tamura20}. 
Following this, we have considered the crucial control problem: how to drive the optical driving system based on the measured distance variation.

\begin{figure} [t!]
   \begin{center}
   \begin{tabular}{c} 
   \includegraphics[width=0.9 \textwidth]{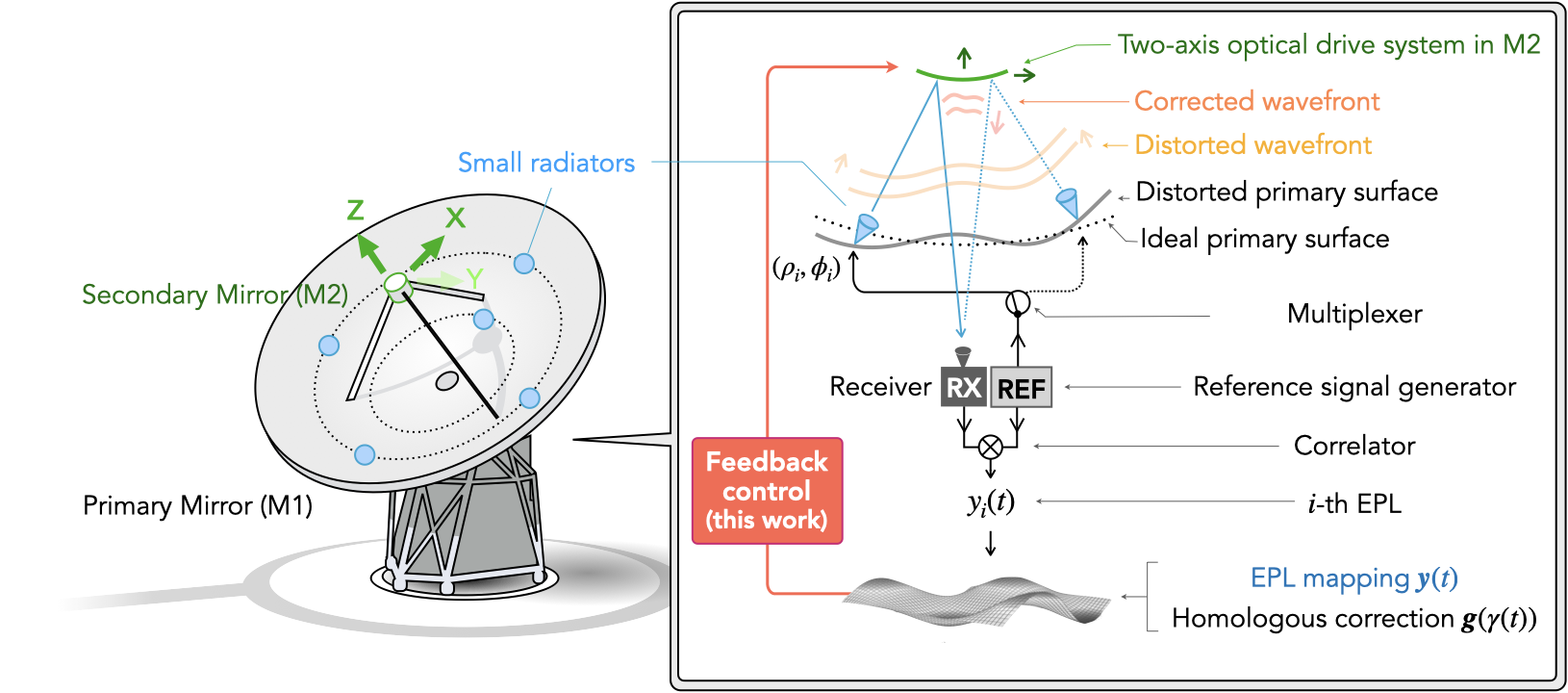}
   \end{tabular}
   \end{center}
   \caption[Concept of MAO]{Conceptual diagram of the Millimeter-wave Adaptive Optics (MAO) system.}
   \label{fig:MAOconcept}
\end{figure}

We formalize this challenge as an asymptotic constant disturbance suppression problem. Considering the physical limitations on the movement of the optical driving system, we have shown that an anti-windup proportional-integral (AWPI) controller provides a robust solution to this problem \cite{Jikuya26a}. 
The positioning of the control law within MAO is illustrated in the conceptual diagram of Fig.~\ref{fig:MAOconcept}.

In this work, we present a simulation result assuming the Nobeyama 45-m Radio Telescope with the sub-reflector (M2) driven by a two-axis translational mechanism. 
For an overview of the experiment at Nobeyama 45-m, please refer to the presentation by Tamura at this conference (Tamura et. al., this conference).
This simulation result demonstrates the effectiveness of the proposed AWPI control approach in accurately compensating for the distance variations. Finally, we discuss future research directions, including implementation considerations and application to other radio astronomy facilities.

\section{Anti-Windup PI Control for MAO}

In this section, we summarize the feedback control framework for MAO by referencing our previous work\cite{Jikuya26a}, in which a general theory applicable to arbitrary MAO configurations is provided. Specifically, we adopt the equations and block diagrams from this general formulation and describe the specific system parameters optimized for its application to the Nobeyama 45-m Radio Telescope.

MAO assumes the use of decoupled multiple subreflector drive systems, where the axes of each drive system are also decoupled. The dynamic characteristics of the subreflector drive system can be expressed by the \textbf{first-order system}, i.e., the system of first-order differential equations: 
\begin{align}
	\bm{T}_{\mathrm{s}} \, \dot{\bm{p}}(t) + \bm{p}(t) = \bm{u}(t). \label{eqn:plant-01}
\end{align}
For the Nobeyama 45-m Radio Telescope, the time constant matrix is given by:
\begin{align}
	\bm{T}_{\mathrm{s}} = 0.2 \bm{I}_2 \in \mathbb{R}^{2 \times 2} \label{eqn:nobT}
\end{align}
where $\bm{T}_{\mathrm{s}}$ is the diagonal matrix with the individual \textbf{time constants} on the diagonal elements, $\bm{I}_2 \in \mathbb{R}^{2 \times 2}$ is the identity matrix of size two, $\bm{p}(t) \in \mathbb{R}^2$ is the column vector of \textbf{drive amounts}, $\bm{u}(t) \in \mathbb{R}^2$ is the column vector of the position commands for the X and Z axes of the subreflector, and the constant $2$ represents the degrees of freedom in the entire optical drive system.

The \textbf{Excess Path Length (EPL) measurements} $\bm{y}(t) \in \mathbb{R}^5$ are obtained in real time, where the constant $5$ represents the degrees of freedom in the EPL measurement system, i.e., the number of radiators. 
The measurements $\bm{y}(t)$ are modeled by the sum of the desired optical path length $\bm{g}(\gamma(t))$, which is associated with the homologous deformation and depends on the time-varying \textbf{elevation angle} $\gamma(t) \in \mathbb{R}$, the optical path length changes $\bm{M} \bm{p}(t) $ at the measurement points, and the \textbf{disturbance} represented by $\bm{w}(t)$ encompassing all unmodeled factors as follows:
\begin{align}
	\bm{y}(t) =  \bm{M} \bm{p}(t) + \bm{g}(\gamma(t)) + \bm{w}(t). \label{eqn:plant-02}
\end{align}
The measurement matrix for the Nobeyama 45-m Radio Telescope is given by:
\begin{align}
	\bm{M} = \begin{bmatrix} 1.960 & -0.259 \\ 1.607 & -0.731 \\ 1.607 & 0.0 \\ 1.607 & 0.731 \\ 1.607 & 0.0 \end{bmatrix} \in \mathbb{R}^{5 \times 2} \label{eqn:nobM}
\end{align}
which is the \textbf{measurement matrix} of column-full rank obtained via electromagnetic field analysis.

The goal of the \textbf{asymptotic disturbance suppression control} is to design a controller mapping from $\bm{y}(t)$ to $\bm{u}(t)$ such that the closed-loop system is asymptotically stable and that the following asymptotic convergence properties are satisfied:
\begin{align}
	\lim_{t \rightarrow \infty} \bm{M}^{\dagger} \bm{q}(t) 	= \bm{0}_{2}
\end{align}
for any initial condition $\bm{p}(0)$ and any constant disturbance $\bm{w}$, 
i.e., the suppressible part of $\bm{w}$ is asymptotically suppressed. 
Here, $\bm{M}^{\dagger} := \left( \bm{M}^{\top} \bm{M} \right)^{-1} \bm{M}^{\top}$ denotes the Moore-Penrose pseudo-inverse matrix, and $\bm{0}$ is the zero vector or matrix with its dimension indicated by the subscript.

\begin{figure} [ht!]
   \begin{center}
   \begin{tabular}{c} 
   \includegraphics[width=0.75 \textwidth]{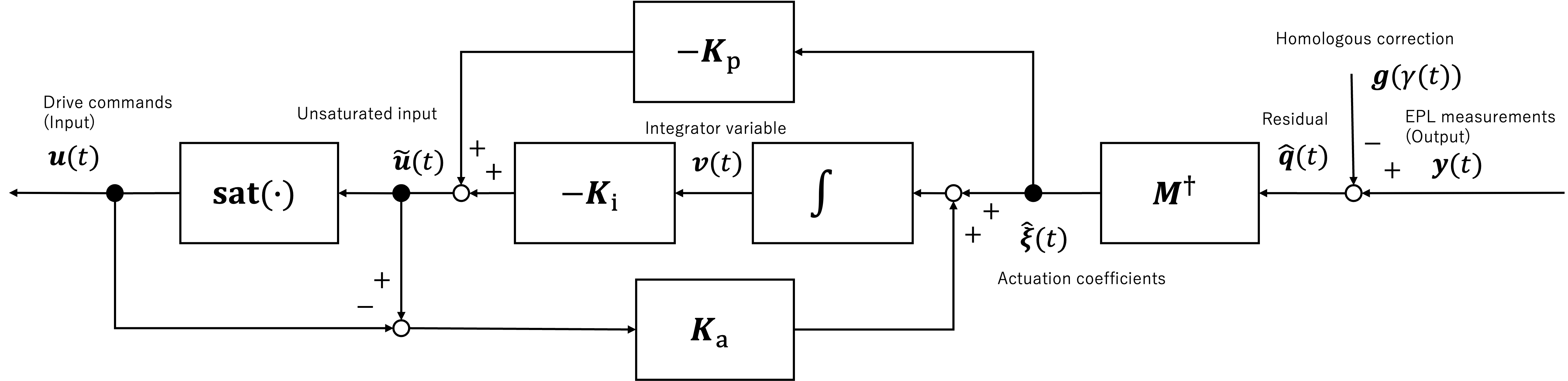}
   \end{tabular}
   \end{center}
   \caption[Block Diagram of AWPI　Controller]{Block diagram of the AWPI controller.}
   \label{fig:AWPIposter}
\end{figure}

This control problem is addressed within the framework of the \textbf{Anti-Windup Proportional-Integral (AWPI) controller} formulated as follows:
\begin{align}
	\hat{\bm{q}}(t) &= \bm{y}(t) - \bm{g}(\gamma(t)), \label{eqn:AWPI-00} \\
	\hat{\bm{\xi}}(t) &= \bm{M}^{\dagger} \hat{\bm{q}}(t), \label{eqn:AWPI-0} \\
	\dot{\bm{v}}(t) &= \hat{\bm{\xi}}(t) + \bm{K}_{\mathrm{a}} \left( \tilde{\bm{u}}(t) - \bm{u}(t)   \right), \label{eqn:AWPI-01} \\
	\tilde{\bm{u}}(t) &= - \bm{K}_{\mathrm{i}} \, \bm{v}(t) - \bm{K}_{\mathrm{p}} \, \hat{\bm{\xi}}(t), \label{eqn:AWPI-02} \\
	\bm{u}(t) &= \mathrm{\mathbf{sat}} \left( \tilde{\bm{u}}(t) \right), \label{eqn:AWPI-03}
\end{align}
where $\hat{\bm{q}}(t)$ is the \textbf{estimated residual}, $\hat{\bm{\xi}}(t)$ is the \textbf{estimated actuation coefficients}, $\bm{v}(t)$ is the \textbf{integrator variable}, $\tilde{\bm{u}}(t)$ is the \textbf{unsaturated input}. 
The input-output relationship of the AWPI control can be visualized using the block diagram shown in Figure~\ref{fig:AWPIposter}.

The matrices $\bm{K}_{\mathrm{i}}$, $\bm{K}_{\mathrm{p}}$, and $\bm{K}_{\mathrm{a}}$ represent the integral gain, proportional gain, and anti-windup gain, respectively. Through loop shaping, they are selected as $\bm{K}_{\mathrm{i}} = k_{\mathrm{i}} \bm{I}_2$ with $k_{\mathrm{i}} = 1.57$ for Nobeyama 45-m Radio Telescope, and $\bm{K}_{\mathrm{p}} = k_{\mathrm{p}} \bm{I}_2$ with $k_{\mathrm{p}} = 0.316$ for Nobeyama 45-m Radio Telescope. The anti-windup gain is chosen as $\bm{K}_{\mathrm{a}} = \bm{K}_{\mathrm{p}}^{-1}$. The function $\bm{\mathrm{sat}}(\cdot)$ represents the saturation function defined below, which restricts each component within the saturation limits $\pm \bar{u}_i$:
\begin{align}
	\mathrm{sat}_i \left( x \right) 
	:= \left\{ \begin{array}{ll} \bar{u}_i & \bar{u}_i < x \\ 
	x & - \bar{u}_i \leq x \leq \bar{u}_i \\ 
	- \bar{u}_i & x < - \bar{u}_i  \end{array} 
	\right. \label{eqn:satdef}
\end{align}
Here, $\tilde{\bm{u}}(t)$ denotes the unsaturated control input, and $\bm{u}(t)$ represents the actual control input considering input saturation. The anti-windup gain matrix $\bm{K}_{\mathrm{a}} \in \mathbb{R}^{2 \times 2}$ is a diagonal matrix configured with positive diagonal elements.

The anti-windup mechanism, consisting of the saturation function and the anti-windup gain, forms a stable local feedback loop. When the control input experiences saturation, this mechanism serves to prevent the divergence of the internal integrator variable $\bm{v}(t)$. Conversely, when the control input desaturates, it effectively suppresses large fluctuations in the control input.

It should be noted that the anti-windup mechanism is strictly a safety feature designed to handle input saturation; under normal operating conditions, the system is expected to operate without reaching saturation. In such unsaturated scenarios, the anti-windup mechanism can be neglected, reducing the closed-loop system to a decoupled first-order system with a standard PI controller. Internal stability of the system is strictly guaranteed as long as both the integral and proportional gains remain positive.

\begin{figure} [t!]
   \begin{center}
   \begin{tabular}{c} 
   \includegraphics[width=0.75 \textwidth]{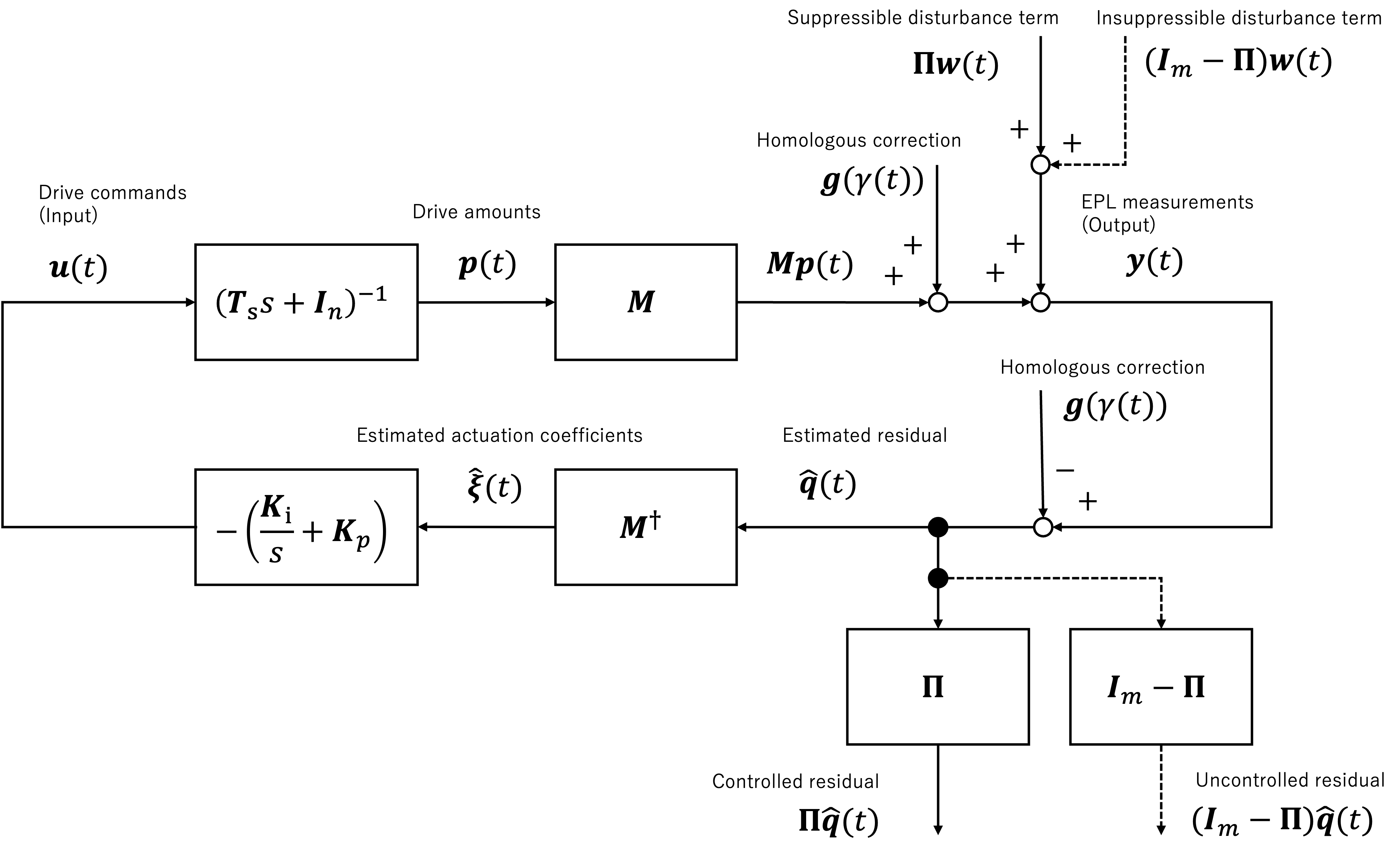}
   \end{tabular}
   \end{center}
   \caption[Block Diagram of Closed Loop System]{Block diagram of the closed-loop system.}
   \label{fig:DirectionalSuppressibility}

   \vspace{1em} 

   \begin{center}
   \begin{tabular}{c} 
   \includegraphics[width=0.75 \textwidth]{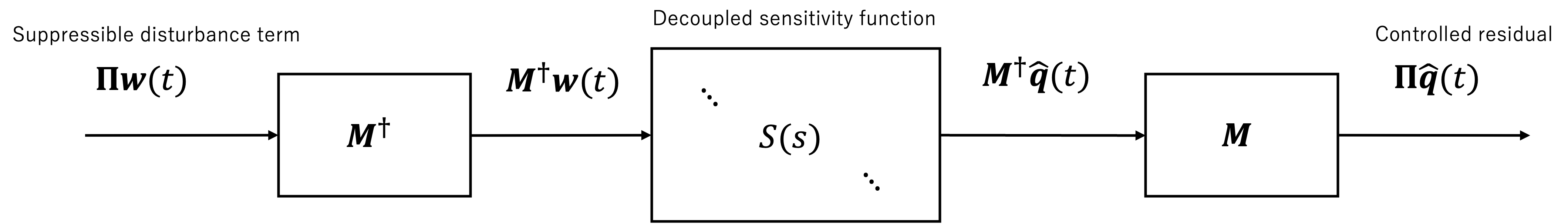}
   \end{tabular}
   \end{center}
   \caption[Decoupled Sensitivity Function]{Structure of the sensitivity function.}
   \label{fig:DecoupledSensitivity}
\end{figure}

The block diagram of the overall closed-loop system is illustrated in Fig.~\ref{fig:DirectionalSuppressibility}. By decomposing the disturbance $\bm{w}$ as follows:
\begin{align}
	\bm{w} = \bm{\Pi} \bm{w} + \left( \bm{I}_5 - \bm{\Pi} \right) \bm{w},
\end{align}
where the first term on the right-hand side represents the suppressible disturbance component, while the second term represents the insuppressible disturbance component. 
Here, $\bm{\Pi} := \bm{M} \bm{M}^{\dagger}$ denotes the orthogonal projection matrix and $\bm{I}_5 \in \mathbb{R}^{5 \times 5}$ is the identity matrix of size five.

The input-output relationship of the suppressible disturbance in the frequency domain is shown in Fig.~\ref{fig:DecoupledSensitivity}.
Based on the modal transformation, the decoupled system behavior is formulated as follows:
\begin{align}
	\bm{\Pi} \bm{\hat{q}}(s) = \bm{M} 
	\mathrm{diag} \begin{bmatrix} \ddots & & \\ & S(s) & \\ & & \ddots \end{bmatrix} 
	\bm{M}^{\dagger} \bm{\Pi} \bm{w}(s).
\end{align}
This architecture structurally places the decoupled sensitivity function $S(s)$ between $\bm{M}$ and $\bm{M}^{\dagger}$. 
In this formulation, the direction of the suppressible disturbance $\bm{\Pi} \bm{w}(s)$ is specified by $\bm{M}^{\dagger}$, while that of the controlled residual $\bm{\Pi} \bm{\hat{q}}(s)$ is determined by $\bm{M}$. 
After reducing the dimensions from an $m$-dimensional signal space to an $n$-dimensional space, the signals are completely decoupled, and the frequency characteristics of this disturbance space are governed by $S(s)$. 
Furthermore, owing to the integrator embedded in the AWPI controller, $S(s)$ functions to effectively reject the disturbance in the low-frequency range.

\section{Simulation Results}

This section presents the simulation results for the Nobeyama 45-m telescope. Specifically, the simulation is performed by applying the AWPI controller described in Eqs.~\eqref{eqn:AWPI-00} to \eqref{eqn:AWPI-03} to the controlled plant defined in Eqs.~\eqref{eqn:plant-01} to \eqref{eqn:nobM}. In accordance with the asymptotic disturbance suppression framework that assumes constant disturbances, we utilize a piecewise constant disturbance function. This allows us to verify both the stability of the closed-loop feedback control system and its disturbance suppression performance across multiple disturbance directions.

The disturbance $\bm{w}(t)$ is formulated as a piecewise constant function as follows:
\begin{align*}
	\bm{w}(t) &:= \left\{ \begin{array}{ll}
	10 \bm{w}_1 & 10 < t < 20  \\
	10 \bm{w}_2 & 30 < t < 40 \\
	20 \bm{w}_1 & 50 < t < 60 \\
	20 \bm{w}_2 & 70 < t < 80 \\ 
	\bm{0} & \mathrm{else} 
	\end{array} 	\right. , \quad
	\bm{w}_1 := \begin{bmatrix} 0.435 \\ 0.302 \\ 0.302 \\ 0.376 \\ 0.698 \end{bmatrix}, \quad 
	\bm{w}_2 := \begin{bmatrix} 0.434 \\ 0.224 \\ 0.224 \\ 0.597 \\ 0.597 \end{bmatrix}.
\end{align*}
Here, the constant vector $\bm{w}_1$ satisfies $\bm{w}_1 \notin \Image \, \bm{M}$, representing a partially suppressible (not completely suppressible but not completely insuppressible) disturbance direction. Conversely, the vector $\bm{w}_2$ satisfies $\bm{w}_2 \in \Image \, \bm{M}$, representing a completely suppressible disturbance direction. To evaluate the system behavior under input constraints, two different amplitudes are configured: an amplitude of $10$ simulates an unsaturated condition, whereas an amplitude of $20$ is intended to induce input saturation. Consequently, simulations are conducted across these four distinct scenarios of constant disturbances.

\begin{figure} [t!]
   \begin{center}
   \begin{tabular}{c} 
   \includegraphics[width=0.9 \textwidth]{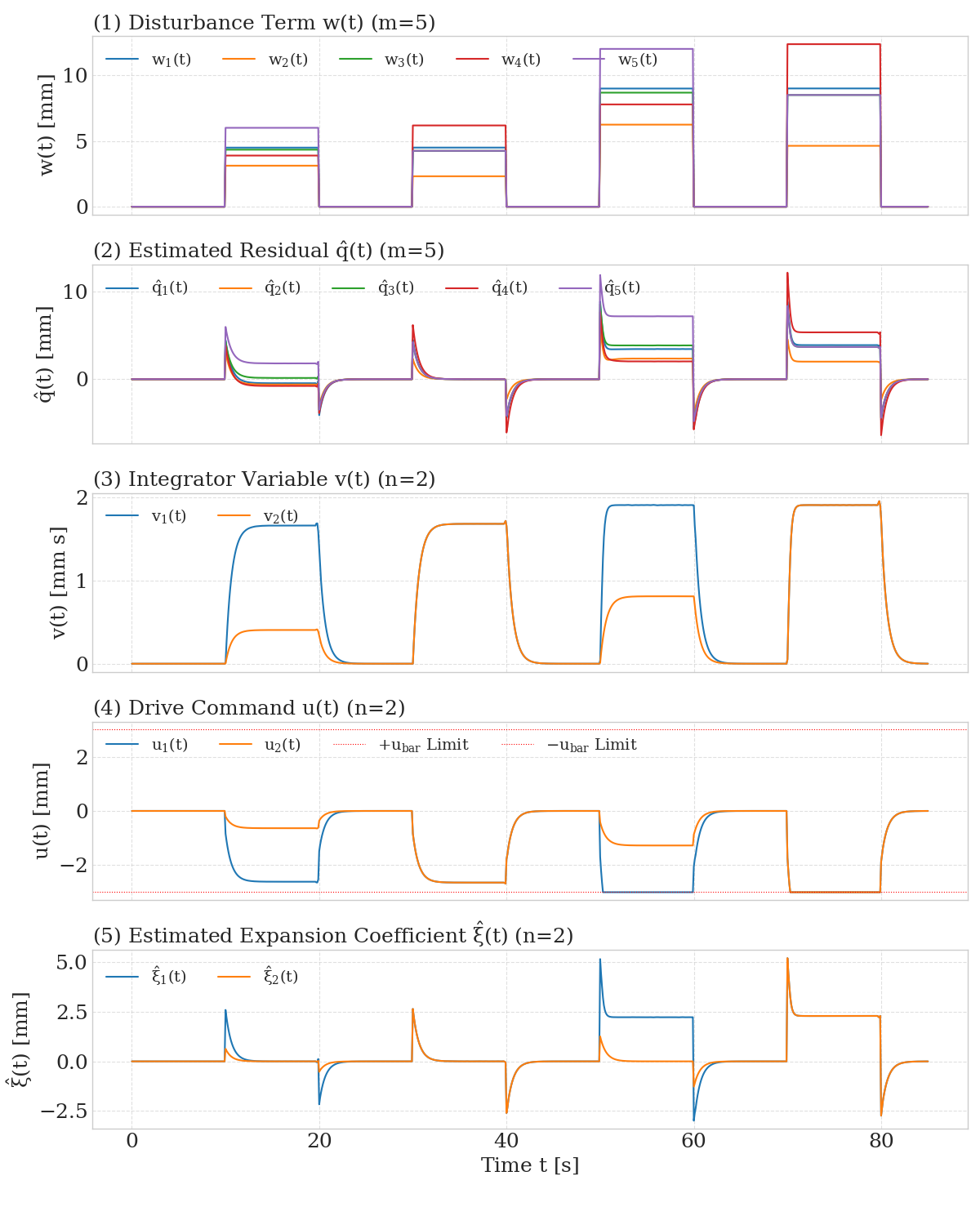}
   \end{tabular}
   \end{center}
   \caption[Simulation Result]{Simulation results of the AWPI controller modeling the Nobeyama 45-m Radio Telescope.}
   \label{fig:FigAWPIsimulation_mscs2026}
\end{figure}

Figure~\ref{fig:FigAWPIsimulation_mscs2026} displays the resulting time histories for the piecewise constant disturbance $\bm{w}$, the residual $\bm{q}(t)$, the internal integrator variable $\bm{v}(t)$, the control input $\bm{u}(t)$, and the estimated expansion coefficients $\hat{\bm{\xi}}(t)$. 
Here, the variables $w_1(t), \dots, w_5(t)$ shown in subfigure (1) represent the individual components of the time-varying vector $\bm{w}(t)$, which should not be confused with the constant vectors such as $\bm{w}_1$ and $\bm{w}_2$. 
In the time interval $10 < t < 20$, the control input does not saturate; although the residual $\bm{q}(t)$ in subfigure (2) does not converge to zero, it successfully stabilizes to a constant value, and the estimated expansion coefficients $\hat{\bm{\xi}}(t)$ in subfigure (5) converge to zero. 
This behavior is reasonable because under the dimensional mismatch ($m =5 > n = 2$), $\hat{\bm{\xi}}(t) \rightarrow 0$ implies that the disturbance components within the suppressible subspace are perfectly rejected, leaving only the insuppressible component in the orthogonal subspace as a constant residual. 
In the interval $30 < t < 40$, where the input remains unsaturated, the residual $\bm{q}(t)$ completely converges to zero. 
During the intervals $50 < t < 60$ and $70 < t < 80$, the control input reaches its saturation limits, as clearly visible in subfigure (4) where $u_1(t)$ hits the lower saturation limit  $- \bar{u}_1$ (indicated by the red dotted line as $- u_{\mathrm{bar}}$). 
Although the estimated expansion coefficients $\hat{\bm{\xi}}(t)$ do not reach zero, they converge to a constant value. 
Crucially, as demonstrated in subfigure (3), the internal integrator variable $\bm{v}(t)$ is prevented from diverging and settles to a stable, flat value, confirming the effective windup protection. 
These simulation results confirm that the closed-loop system performs exactly as designed. 
Particularly in the intervals without input saturation, the asymptotic constant disturbance suppression capability is verified by the convergence of the expansion coefficients $\hat{\bm{\xi}}(t)$ to zero.

In our poster presentation, to complement these findings and demonstrate that the PI controller effectively rejects disturbances at low frequencies via the decoupled sensitivity function $S(s)$, we evaluate the frequency dependence of the suppression performance against suppressible disturbances using the von Karman disturbance model. Since these frequency-domain characteristics are inherently consistent with the results documented in our previous work\cite{Jikuya26a}, readers are referred to that literature for further technical details.

\clearpage

\section{Conclusion}

This paper focused on the control problem in MAO, presenting the underlying theory and corresponding simulation results that form the baseline for control experiments at the Nobeyama 45-m Radio Telescope.
Specifically, using the measurement matrix derived from the 2-degree-of-freedom drive mechanism and the 5-point radiator configuration of the Nobeyama 45-m Radio Telescope, we constructed an AWPI controller. The results verified that the feedback control system operates stably and exhibits the disturbance suppression performance predicted by the theory.

The outcomes of this research are expected to serve as a technical foundation for future large-aperture telescope projects, such as AtLAST and LST. Moving forward, we intend to further refine the control laws based on the AWPI framework and validate their practical effectiveness through field experiments at the Nobeyama 45-m Radio Telescope.

\bibliography{report} 

@INPROCEEDINGS{Tamura20,
       author = {{Tamura}, Yoichi and {Kawabe}, Ryohei and {Fukasaku}, Yuhei and {Kimura}, Kimihiro and {Ueda}, Tetsutaro and {Taniguchi}, Akio and {Okada}, Nozomi and {Ogawa}, Hideo and {Hashimoto}, Ikumi and {Minamidani}, Tetsuhiro and {Kawaguchi}, Noriyuki and {Kuno}, Nario and {Togami}, Yohei and {Hagimoto}, Masato and {Nakano}, Satoya and {Matsuda}, Keiichi and {Okumura}, Sachiko and {Nakamura}, Tomoko and {Kurita}, Mikio and {Takekoshi}, Tatsuya and {Oshima}, Tai and {Onishi}, Toshikazu and {Kohno}, Kotaro},
        title = "{Wavefront sensor for millimeter/submillimeter-wave adaptive optics based on aperture-plane interferometry}",
    booktitle = {Ground-based and Airborne Telescopes VIII},
         year = 2020,
       editor = {{Marshall}, Heather K. and {Spyromilio}, Jason and {Usuda}, Tomonori},
       series = {Society of Photo-Optical Instrumentation Engineers (SPIE) Conference Series},
       volume = {11445},
        month = dec,
          eid = {114451N},
        pages = {114451N},
          doi = {10.1117/12.2561885},
archivePrefix = {arXiv},
       eprint = {2102.09286},
 primaryClass = {astro-ph.IM},
       adsurl = {https://ui.adsabs.harvard.edu/abs/2020SPIE11445E..1NT}
}

@article{Mroczkowski2025,
  author        = {Mroczkowski, Tony and Gallardo, Patricio A. and Timpe, Martin and Kiselev, Aleksej and Groh, Manuel and Kaercher, Hans and Reichert, Matthias and Cicone, Claudia and Puddu, Roberto and {Dubois-dit-Bonclaude}, Pierre and Bok, Daniel and Dahl, Erik and Macintosh, Mike and Dicker, Simon and Viole, Isabelle and Sartori, Sabrina and {Valenzuela Venegas}, Guillermo Andr{\'e}s and Zeyringer, Marianne and Niemack, Michael and Poppi, Sergio and Olguin, Rodrigo and Hatziminaoglou, Evanthia and De Breuck, Carlos and Klaassen, Pamela and Montenegro-Montes, Francisco Miguel and Zimmerer, Thomas},
  title         = {Design of the 50-meter {Atacama} {Large} {Aperture} {Submm} {Telescope}},
  journal       = {Astronomy \& Astrophysics},
  volume        = {694},
  pages         = {A142},
  year          = {2025},
  month         = feb,
  doi           = {10.1051/0004-6361/202418645},
  eprint        = {2402.18645},
  archiveprefix = {arXiv},
  primaryclass  = {astro-ph.IM}
}

@ARTICLE{Jikuya26a,
       author = {{Jikuya}, Ichiro and {Tamura}, Yoichi},
        title = "{Control problem in millimeter-wave adaptive optics}",
      journal = {Journal of Astronomical Telescopes, Instruments, and Systems},
         year = 2026,
        month = apr,
       volume = {12},
          eid = {029005},
        pages = {029005},
          doi = {10.1117/1.JATIS.12.2.029005},
archivePrefix = {arXiv},
       eprint = {2606.09515},
 primaryClass = {astro-ph.IM},
       adsurl = {https://ui.adsabs.harvard.edu/abs/2026JATIS..12b9005J}
}
\bibliographystyle{spiebib} 

\end{document}